\documentclass[a4paper]{VisionStyle}
\usepackage{graphics,epsfig}

\def\ein{{\it Einstein}}
\def\chandra{{\it Chandra}}
\def\xmm{{XMM-Newton}}

\def\ergsec{\hbox{erg s$^{-1}$}}
\def\ergcm{\hbox{erg cm$^{-2}$ s$^{-1}$}}

\def\cm-2{cm$^{-2}$}

\def\HI{\hbox{H\,{\sc i}}}
\def\HII{\hbox{H\,{\sc ii}}}

\def\ngc3079{{\object{NGC~3079}}}
\def\n253{{\object{NGC~253}}}
\def\smc{{\object{SMC}}}
\def\lmc{{\object{LMC}}}
\def\m33{{\object{M33}}}
\def\andro{{\object{M31}}}
\def\ng300{{\object{NGC~300}}}

\begin{document}

\title{XMM-Newton Studies of the Source Population and the Hot Interstellar
Medium in Nearby Galaxies}

\author{W.\,Pietsch\inst{}} 

\institute{Max-Planck-Institut f\"ur extraterrestrische Physik, Postfach 1312,
85741 Garching, Germany}

\maketitle 

\begin{abstract}

First results of X-ray source population studies in \linebreak[4] nearby
galaxies show the potential of \xmm\ observations. 
I will 
report on first \xmm\ \andro\ results and on three of our \xmm\ projects, an 
X-ray source population
study in the Magellanic Clouds (MCs), a deep raster survey of \m33, and an
investigation of the hot interstellar medium (ISM) in the halo of 
edge-on galaxies. \xmm\ results on several other galaxies and sources 
within are presented by other authors in these proceedings.

Our MC study is build up of deep pointings probing MC sources down to 
$10^{33}$~erg~s$^{-1}$\ 
and shallower pointings to confirm candidates from our ROSAT derived
lists of X-ray binaries (XRBs), super-soft sources (SSSs), and 
supernova remnants (SNRs). First
\xmm\ detections of a 455 s pulsar in the Small Magellanic Cloud (\smc)  
and the results of the Large Magellanic Cloud (\lmc)
deep field confirm the validity of our strategy.

Our \m33\ raster pointing aims for luminosities as low as 
$10^{35}$~erg~s$^{-1}$, a
factor of 10 below the sensitivity limit of the ROSAT observations. The survey 
will allow us to
characterize the sources using extent, spectra, hardness ratios and time
variability to build  up a  unprecedented census of the X-ray source content of
\m33. Of specific interest are the active source in the nuclear area and the
diffuse emission in the inner disk.

\xmm\ observations of the active galaxy 
\linebreak[4] 
\ngc3079 and of the starburst galaxy  \n253
are used to characterize the point-like sources and the hot ISM in the 
disk and from the halo of these galaxies.

\end{abstract}

\section{Introduction}
Observations of the \ein\ and ROSAT observatories
demonstrated that the X-ray emission of nearby late type galaxies (spirals and 
irregulars) can be separated
in contributions from point-like sources (low- and high-mass XRBs, SSS, 
young supernovae, SNRs, etc.) and diffuse emission (hot ISM within the disk, 
from
an outflow, or from the galaxy halo). In some of the galaxies X-ray
emission from an active nucleus (AGN) could be detected (sometimes heavily
absorbed) dominating the galaxy's X-ray emission. The imaging proportional counters
aboard \ein\ (IPC) and ROSAT (PSPC) were able to spatially resolve these components. However,
the energy resolution was only sparse. The CCD detectors aboard the ASCA and
BeppoSAX observatories on the other hand provided much better energy resolution
and a broader energy coverage. However with the poorer spatial resolution, in 
most cases it was not possible to resolve the different emission components.
The average parameters extracted for the galaxies as a whole, are of rather limited value.

With the new generation X-ray observatories \chandra\ and \xmm, we now have the 
sensitivity and
spatial and spectral resolution to probe significantly deeper. The high
throughput of \xmm\ combined with the  spatial resolution comparable to the ROSAT
high resolution imager (HRI) allows us in X-ray source population studies 
(1) to detect sources more than a factor of 10 deeper than before;
(2) to classify the sources in a broad X-ray energy band (spectra, hardness
ratios, variability, extent, pulsations);
(3) to determine  log N / log S distributions.
For the hot ISM, we are able
(1) to resolve it from X-ray point sources;
(2) to analyze the spatial variability of its X-ray spectrum with medium and 
high energy resolution. With the results for different galaxies, we can search
for 
dependencies of the X-ray parameters on galaxy type, metallicity, and star 
forming history. 

In the following I will report on \xmm\ results that demonstrate these
possibilities. Interesting additional new \xmm\ results on \andro, \ng300\ 
and other nearby galaxies (and individual sources within) will be presented in 
these proceedings.

\section{First results on \andro}
The Andromeda Galaxy (\andro) is the closest big spiral galaxy to our own and 
was 
observed as a \xmm\ performance verification (PV) target in order to demonstrate 
the
capabilities of the mission in performing spectral and timing studies in a field
of point sources and extended emission. 

In a first paper 
\cite*{wpietsch-E3:shi01} focused on the group properties of the X-ray point
sources and on the diffuse emission. They detect 116 sources in the central
30\arcmin\ of \andro\ down to a limiting luminosity of $6\times10^{35}$ \ergsec.
The luminosity distribution of the sources flattens at luminosities below 
$\sim2.5\times10^{37}$ \ergsec. The detected sources can be classified into
source classes such as SSSs and intrinsically hard or soft sources using 
hardness ratios. The spectrum of the unresolved emission 
in the bulge of \andro\ is demonstrated to contain a soft excess. It can be 
fitted with a 0.35
keV optically-thin thermal-plasma component from diffuse hot gas in the center 
of \andro\ which is clearly distinct from the composite point-source spectrum. 
The extended emission can clearly be seen in the image.

In a second paper
\cite*{wpietsch-E3:osb01} discussed the variability of individual sources in
\andro\ based on PV observations in June and December 2000. 
For the $\sim60$ brightest sources they searched for time variability.
At least 15\% of the sources appear to be variable on a time scale of several
months. A bright new transient source detected close to the nucleus in June, was
no longer detectable in the December observation and may be similar to some 
Galactic low mass XRBs, most of which are supposed to harbour black holes. They
also detected a 865 s period from a SSS, which was significantly fainter
in the second observation. This source may be a rapidly spinning magnetized
white dwarf in a symbiotic nova system.

\section{Magellanic Cloud studies}
The `first light' target of \xmm\ was the X-ray rich \lmc\ region close to 30 
Doradus and supernova \object{SN1987A} (\cite{wpietsch-E3:den01}). The X-ray 
spectrum of several SNRs in the field was analyzed using the EPIC cameras. The 
hot ISM in the area indicates significant overabundance of Ne and Mg. Column 
densities derived from the EPIC spectra of several AGN shining through the 
interstellar gas of the \lmc\ compared with \HI\ measurements, will allow to draw 
conclusions on \lmc\ atomic to molecular hydrogen ratios 
(\cite{wpietsch-E3:ha101}).

Several more \lmc\ and \smc\ sources were observed as calibration and PV targets 
and characterized by high resolution RGS or EPIC spectra. Examples are the SNRs 
\linebreak[4]
\object{1E 0102.2-7219} (\cite{wpietsch-E3:ras01}, \cite{wpietsch-E3:sa101}), 
\object{N132D} (\cite{wpietsch-E3:beh01}), \object{SNR B0540-69.3} 
(\cite{wpietsch-E3:van01}), and the SSS \object{CAL 83} 
(\cite{wpietsch-E3:pae01}). For the candidate black hole XRB \object{LMC X-3}, 
EPIC and RGS observations cover the soft and hard spectral state 
(\cite{wpietsch-E3:wu01}). OM filter observations let to a more accurate 
lower mass limit of the compact object in  \object{LMC X-3} 
(\cite{wpietsch-E3:sor01}).

We analyzed in detail of all ROSAT PSPC and HRI observations of 
the MC which led to large catalogues of X-ray sources
(\lmc: \cite{wpietsch-E3:hab99}, \cite{wpietsch-E3:sas00}; 
\smc: \cite{wpietsch-E3:hab00}, \cite{wpietsch-E3:sa100})
which we characterized by their hardness ratios and extent and partly 
identified by cross-correlations with catalogues at other wavelengths.
To follow up on this work, 
we proposed deep field observations in some areas and shallower observations 
of some XRB and SNR candidates to search for periods and/or measure extent and
spectra. 

Haberl (see these proceedings) reports on an \lmc\ deep guaranteed time (GT)
pointing to a field containing three high mass XRBs which were 
identified in ROSAT studies
during outburst. The XRBs are detected with the \xmm\ EPIC instruments 
in quiescence and still periods can be derived.
In a preliminary overall analysis of the field, 
about 150 X-ray sources are detected down to fluxes of 
$10^{-14}$~\ergcm,  while ROSAT only detected 35 sources in the field. 
New faint SNRs and XRBs are detected and also a class of new faint sources that most
likely represent low mass XRBs and/or cataclysmic variables in the \lmc\ and 
are not background AGN.

\begin{figure}[t]
  \begin{center}
    \includegraphics[bb=12 14 307 310,width=8.9cm,clip=]{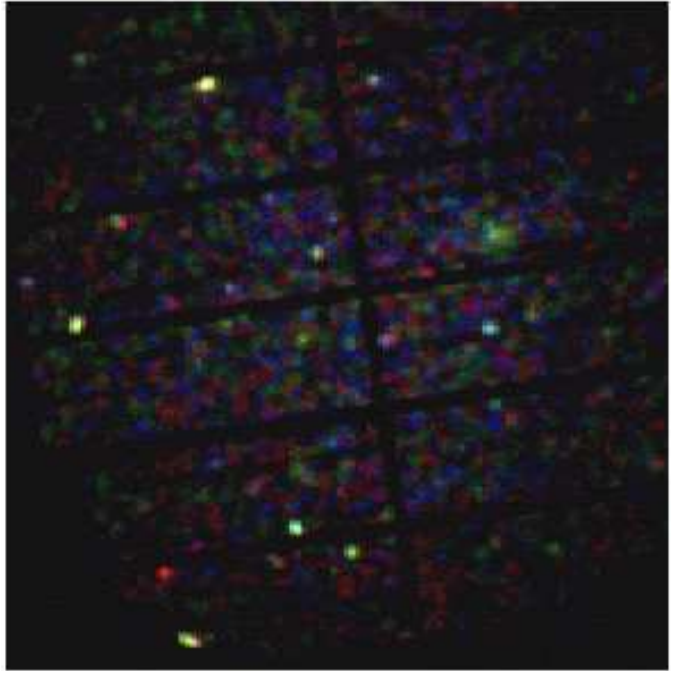}
  \end{center}
\caption{ EPIC PN image of a $\sim13$\arcmin\ radius region around
\object{SMC X-2}. The colours represent X-ray intensities in different energy
bands (red: 0.2--1.0 keV, green: 1.0--2.0 keV, blue: 2.0--5 keV). Red objects
are foreground stars. The other objects are high mass XRB candidates or
background active nuclei. The transient pulsar \object{SMC X-2} was detected in
quiescence (blue object in the center). The extended greenish object in the NW
is most likely a new \object{SMC} SNR. }  
\label{wpietsch-E3_fig1}
\end{figure}

During October 2001, a series of \xmm\ pointings was
started to further constrain our list of ROSAT \smc\ XRB candidates and
detect new XRBs and XRB pulsars in the \smc. 
The first target was the area around the Be/XRB transient 
\object{SMC X-2} (see Fig.~\ref{wpietsch-E3_fig1}). Unfortunately,
the observation suffered from high EPIC background. Nevertheless,
we did not only detect \object{SMC X-2} in quiescence at a luminosity as low as 
$1.5\times10^{33}$~\ergsec, but besides foreground stars, also a new SNR 
and high mass XRB candidates (from comparison with digital sky survey images).

\begin{figure}[t]
 \begin{center}
    \includegraphics[width=3.0cm,angle=-90,bb=65 40 529 700,clip=]{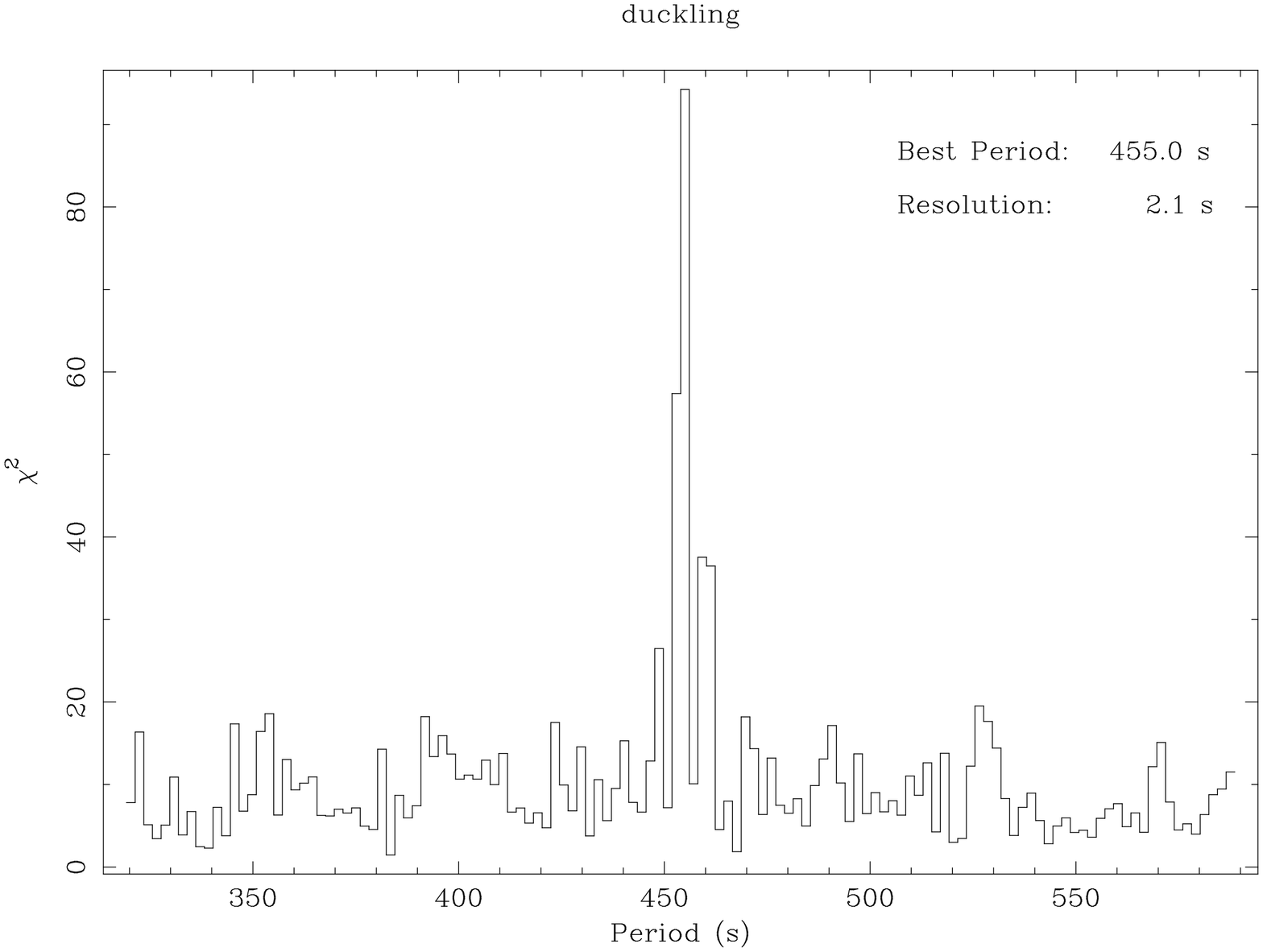}
    \includegraphics[width=3.0cm,angle=-90,bb=65 40 529 700,clip=]{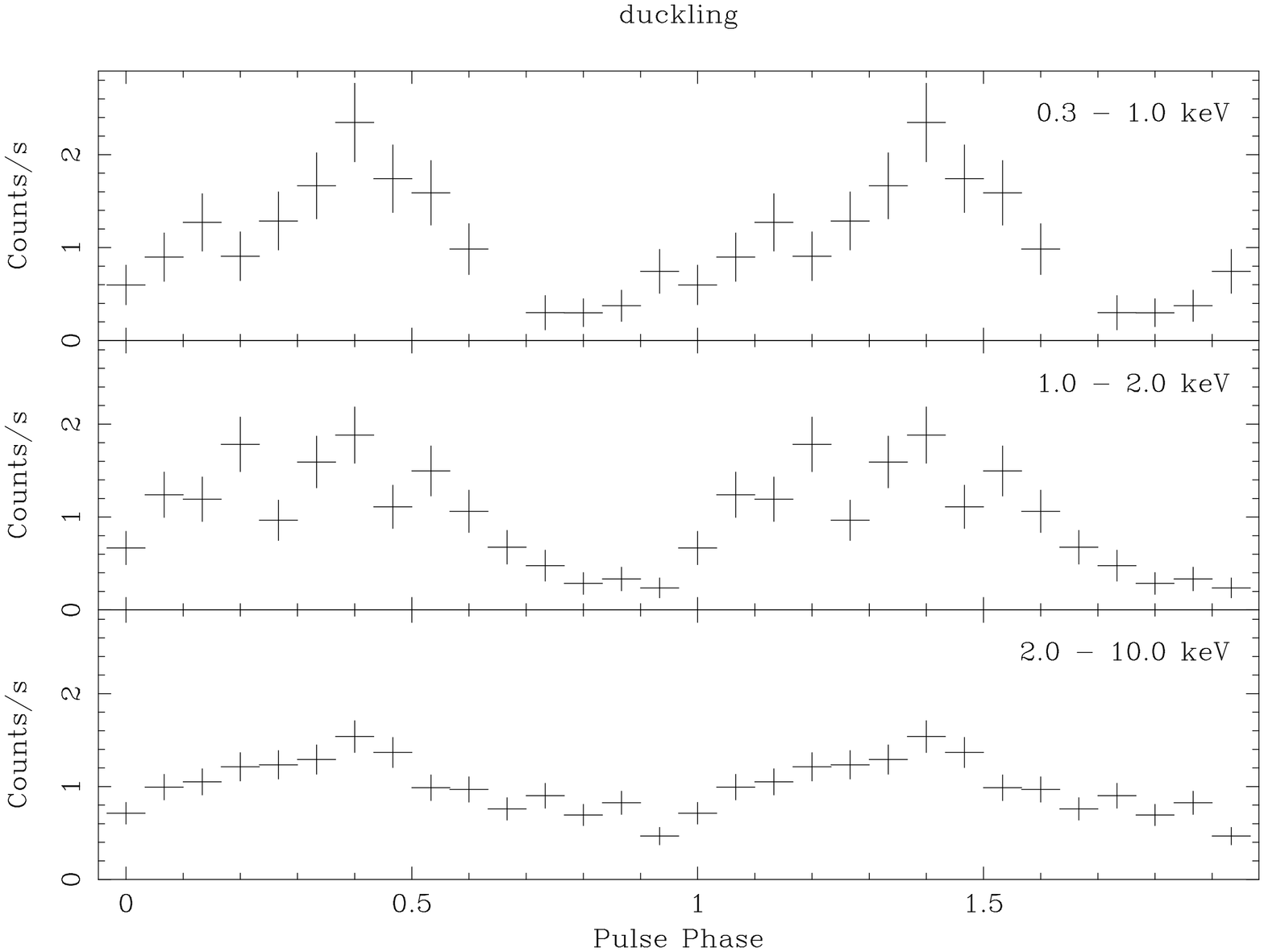}
  \end{center}
\caption{ XMM-Newton EPIC PN detection of 455 s pulsation from the Be/XRB 
\object{RX J0101.3-7211}: ({\bf Left}) $\chi^2$ periodogram, showing the strong 
signal at the
proposed period.({\bf Right}) light curve folded modulo 455 s for 3 energy 
bands.}  
\label{wpietsch-E3_fig2}
\end{figure}

We also tried to use calibration and PV observations to prove our search
strategy for XRB pulsars. We serendipitously detected the ROSAT 
Be/XRB candidate \linebreak[4] 
\object{RX J0101.3-7211} 
in the EPIC PN instrument at the outer boarder of a \object{SNR 0102-72.3} 
observation (\cite{wpietsch-E3:sas01}). The luminosity of 
\object{RX J0101.3-7211} determined during the observation with the \xmm\ 
EPIC PN in the ROSAT band ($2.5\times10^{34}$~\ergsec), was lower than during the 
faintest ROSAT detection; the X-ray spectrum
can be described by a power-law with photon index of 0.6.  
A timing analysis of the data 
revealed 455 s pulsations (Fig.~\ref{wpietsch-E3_fig2}) that are clearly
detectable in the energy bands 0.3--1.0, 1.0--2.0, and 2.0--10.0 keV with a
slight shift of the minimum to earlier phases in the softest band. 
These findings together
with optical spectra of the proposed counterpart confirm 
\object{RX J0101.3-7211} as \smc\ Be/XRB.

\section{XMM-Newton \object{M33} raster observation}
Besides \andro\ and the MCs the Sc spiral \m33\ seen near face-on, belongs to 
the galaxies in the Local Group best studied in X-rays. 
We carefully analyzed all \m33\ ROSAT observations in the archive and found 
184 X-ray sources within 50\arcmin\ of
the nucleus. We partly classified and identified the 
sources by correlations with previous catalogues and using their X-ray properties.
\m33\ is known to harbour a bright active source in the nuclear area, several
SSSs, SNRs, and XRB candidates (see \cite{wpietsch-E3:hab01} and references therein). 
For one of the sources, eclipses with a period of
3.45 d and a possible 0.31 s pulsation period strongly suggest a high mass XRB
nature (\cite{wpietsch-E3:dub99}).

As a GT program we proposed fifteen 10 ks pointings to homogeneously
cover \m33\ to unprecedented depth (better than $10^{35}$~\ergsec\ within
the optical $D_{25}$ diameter) and smooth out residual EPIC detector 
structures by this raster. The first five observations were performed in August 2000
and confirmed our strategy (see \cite{wpietsch-E3:pie00} and
\cite{wpietsch-E3:ehl01}).

\begin{figure}[ht]
  \begin{center}
    \includegraphics[bb=37 120 557 725,width=8.9cm,clip=]{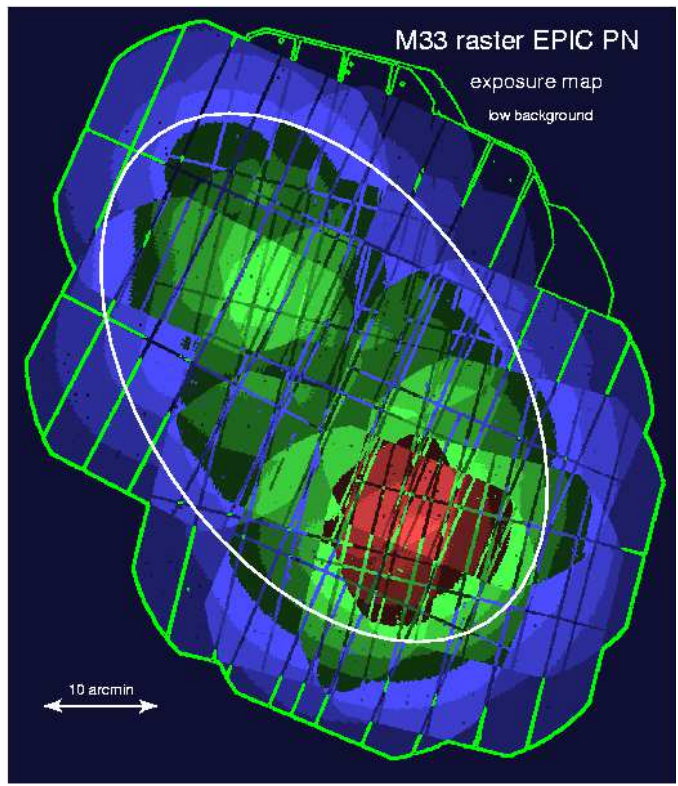}
  \end{center}
\caption{ Exposure map of the EPIC PN low background times for the 12 M33 raster
observations. Blue colours give exposures from 0--10 ks, green from 10--20 ks
and red from 20--30 ks. The white ellipse indicates the optical $D_{25}$ 
diameter
of the galaxy.}  
\label{wpietsch-E3_fig3}
\end{figure}

Until August 2001, seven more pointings of the raster were performed.
Unfortunately, one of the observations was rather short and several others 
suffer from
high detector background. To produce homogeneous images and to reach highest
sensitivity for faint sources and diffuse emission from \m33, we had to reject 
high background times. The resulting combined low background EPIC PN 
exposure map for the 12 raster observations is shown in 
Fig.~\ref{wpietsch-E3_fig3}. While in the SW part of
the \m33\ disk we achieve an exposure of $\sim$30 ks, the central and  NE part is
much less exposed.

\begin{figure}[ht]
  \begin{center}
    \includegraphics[width=8.9cm,bb=79 146 533 646,clip=]{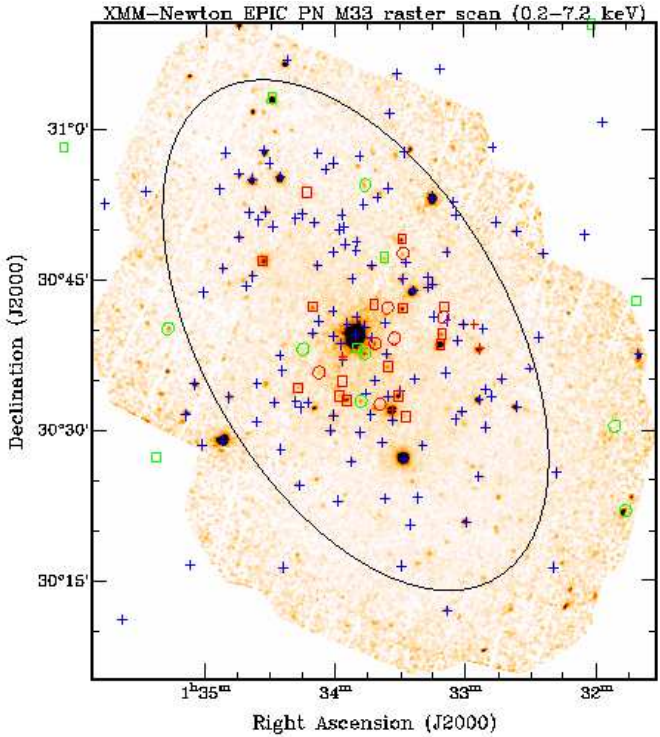}
  \end{center}
\caption{EPIC PN low background image of the 12 M33 raster
observations (0.2--7.2 keV). The black ellipse indicates the 
optical $D_{25}$ diameter
of the galaxy. ROSAT sources from \protect\cite*{wpietsch-E3:hab01} are 
superimposed. 
Classified sources are indicated: SNRs by red squares, SSS by red circles, 
XRBs by red crosses,
foreground stars by green circles, background AGN by green squares. 
Blue crosses indicate unclassified sources.}  
\label{wpietsch-E3_fig4}
\end{figure}

\begin{figure}[ht]
  \begin{center}
    \includegraphics[bb=17 20 533 622,width=8.9cm,clip=]{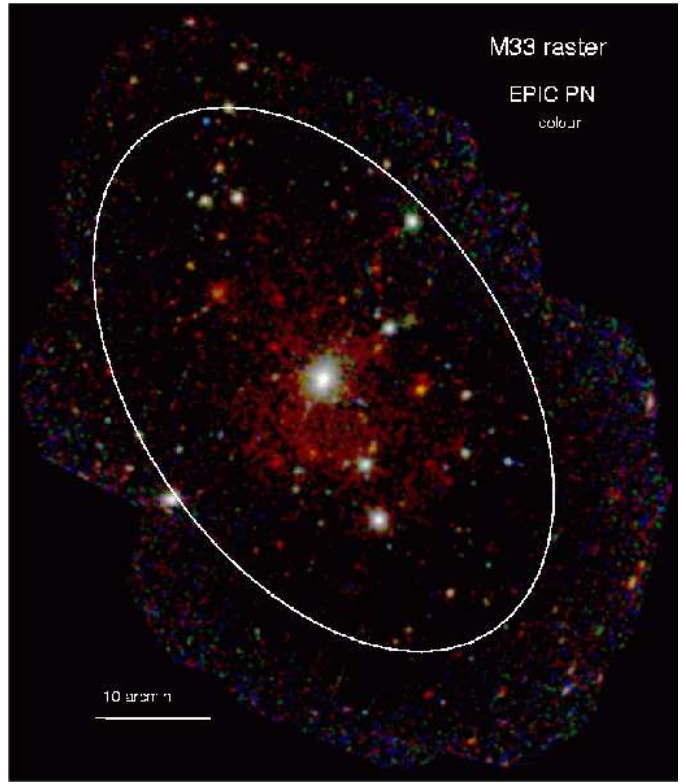}
  \end{center}
\caption{EPIC PN low background image of the 12 M33 raster
observations. The colours represent X-ray intensities in different energy
bands (red: 0.2--1.0 keV, green: 1.0--2.0 keV, blue: 2.0--7.2 keV).
The white ellipse indicates the optical $D_{25}$ diameter
of the galaxy. }  
\label{wpietsch-E3_fig5}
\end{figure}

Figure \ref{wpietsch-E3_fig4} gives a slightly smoothed
EPIC PN low background image of the \m33\ raster in the 0.2--7.2 keV band, 
Fig.~\ref{wpietsch-E3_fig5} the corresponding colour image (red: 0.2--1.0 keV,
green: 1.0--2.0 keV, blue: 2.0--7.2 keV). The optical 
$D_{25}$ diameter is indicated. In Fig.~\ref{wpietsch-E3_fig4}, ROSAT sources 
from \cite*{wpietsch-E3:hab01} are
superimposed. To create the images we used photon event lists with sky coordinates 
transformed to a common reference point.
For each observation we separately created low background images for the 
0.2--0.5, 0.5--1.0, 1.0--2.0, and 2.0--7.2 keV bands and accepted
just single events for the lowest energy band, singles and doubles for the 
other bands. We subtracted `out of time event' images. We also created 
individual exposure maps.
The resulting individual images and exposure maps were masked for low
exposure and then combined. The combined images were exposure corrected and 
finally the images for the bands were added as needed.  

A large number of point-like sources (several of which were identified by ROSAT
before, see \cite{wpietsch-E3:hab01}), some diffuse sources and strong emission
from the nuclear area are clearly visible. Diffuse emission is detected
surrounding the bright nucleus and from the location of the optically bright
southern spiral arm. The colour image shows SSS in red, SNRs in yellow, and
XRBs as well as background AGN in white or blue. The extended red object NE of the
nucleus is the giant \HII\ region \object{NGC 604}. 

To better characterize the sources, we performed spectral fits to the EPIC PN
data of the brightest sources and produced light curves. EPIC MOS data will be
added. Some first results are given below.

The nuclear source is
the most luminous persistent source in the Local Group (unabsorbed 0.3-10 keV
luminosity of $\sim1.2\times10^{39}$~\ergsec). Its EPIC PN spectrum can
be described by a disk-blackbody plus power law model typical for black hole
XRBs. The best fit parameters during the pointing when the source was in the
center of the field, are: $N_{\rm H} = 2.1\times10^{22}$ \cm-2, 
well above the Galactic value, photon index 2.6, temperature 
$kT_{\rm in} = 1.2$ keV at the inner disk radius, 
$R_{\rm in}{\rm cos}(i)^{0.5} = 31$~km, yielding an estimated Schwarzschild
black hole mass of $\sim3.5$cos$(i)^{-0.5}$~M$_{\sun}$. No time variability was
detected during this observation. RGS spectra are in good agreement with the
EPIC PN results and do not show any significant emission or absorption lines.

Several XRB and SSS candidates show variability with\-in individual observations.
The \cite*{wpietsch-E3:dub99} high mass XRB was in the field of view during
seven pointings. In two observations at extrapolated binary phases 0.0 and 0.1, 
it was faint, while during the rest of the observations 
(at phases 0.2 to 0.6), it
showed no variability. A background AGN candidate showed an absorbed  power 
law spectrum with a photon index of 2.0.

\section{Hot ISM in edge-on galaxies} 
ROSAT has detected diffuse emission from the hot ISM in the disk, from
outflowing winds (e.g. the `plume' in \n253 extending more than 1\arcmin\ to
the SE from the nucleus along the minor axis), 
and from the halos of several edge-on galaxies. Many of them
were known for star-forming or nuclear activity. Several of these galaxies have 
been or will be observed by \xmm\ to constrain proposed equilibrium and non 
equilibrium models with better spectra and specifically to investigate if the
X-ray spectra really point at strong subsolar metallicity in these plasma. 
 
\begin{figure}[t]
  \begin{center}
    \includegraphics[bb=125 219 483 572,width=8.9cm,clip=]{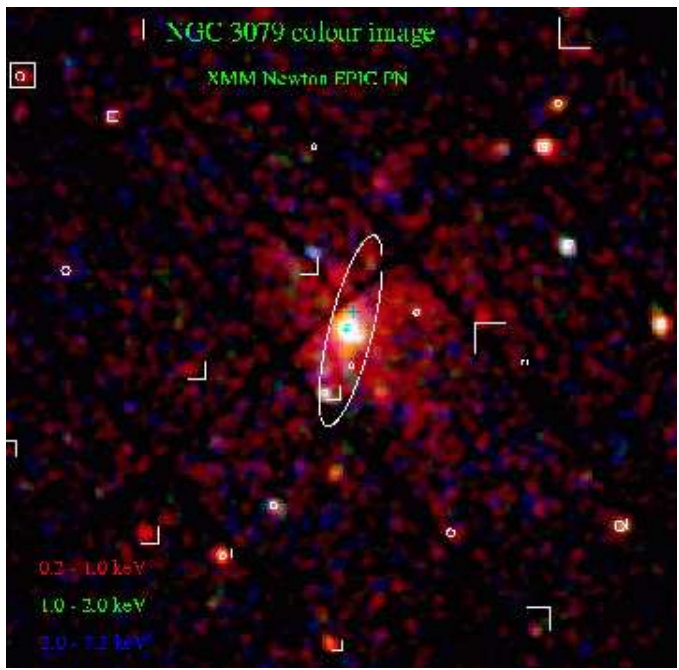}
  \end{center}
\caption{EPIC PN low background image for the \ngc3079\ field. 
The colours represent X-ray intensities in different energy
bands (red: 0.2--1.0 keV, green: 1.0--2.0 keV, blue: 2.0--7.2 keV).
The white ellipse indicates the optical $D_{25}$ diameter
of the galaxy. ROSAT PSPC sources from  \protect\cite*{wpietsch-E3:pie98} are 
indicated by squares, HRI sources by
circles. The symbol size represents the 90\% confidence positional error. The
position of SN 2001ci is indicated by the blue cross north of the galaxy 
center.}  
\label{wpietsch-E3_fig6}
\end{figure}

We have analyzed PV observations on \n253\ and GT observations on \n253,
\object{NGC 2146}, and \ngc3079. Unfortunately, the EPIC PN thin filter observations on
these galaxies were hampered by strong background flaring. Here I 
will present a first EPIC PN colour image of \ngc3079\ and then give more 
details on \n253.

\ngc3079\ was observed by \xmm\ in April 2001. The EPIC PN colour image 
(see Fig.~\ref{wpietsch-E3_fig6}) shows
besides point sources in the field, sources within the optical $D_{25}$ diameter 
of the galaxy that coincide with the ROSAT detected sources 
(\cite{wpietsch-E3:pie98}). The nuclear area is very bright and shows a hard
spectrum. The hard source coincides with the radio nucleus whereas the softer
bright nuclear emission is slightly offset to the E and originates from the 
nuclear outflow in this direction. The extended diffuse emission from the outer
disk and halo of \ngc3079\ is mainly
detected below 1 keV (red) and shows the morphology known from the ROSAT PSPC
images. We do not see any source at the position of \object{SN 2001ci}, a type
Ic supernova that exploded in \ngc3079\ just four to twelve days after the \xmm\
observation.

The PV observation of \n253\ in July 2000 were split for EPIC PN in a 39 ks
and a 14 ks exposure with medium and thin filter, respectively. 
While the first exposure mainly showed
low background, half of the thin filter exposure had to be rejected due to
background flaring. For a first analysis of the PV results see 
\cite*{wpietsch-E3:pie01}. Our 24.5 ks GT observation of \n253\ in December 2000
was pointed slightly offset from the nucleus into the NW halo. EPIC
PN was used with the thin filter  during the observation. Here we present 
some first results from this observation just for EPIC PN. 
Only 8.3 ks of the observation could be collected with 
low background while for the rest of the time the background was higher by a 
factor of at least two or even flaring. 

\begin{figure*}[ht]
 \begin{center}
    \includegraphics[width=8.9cm,bb=130 241 482 548,clip=]{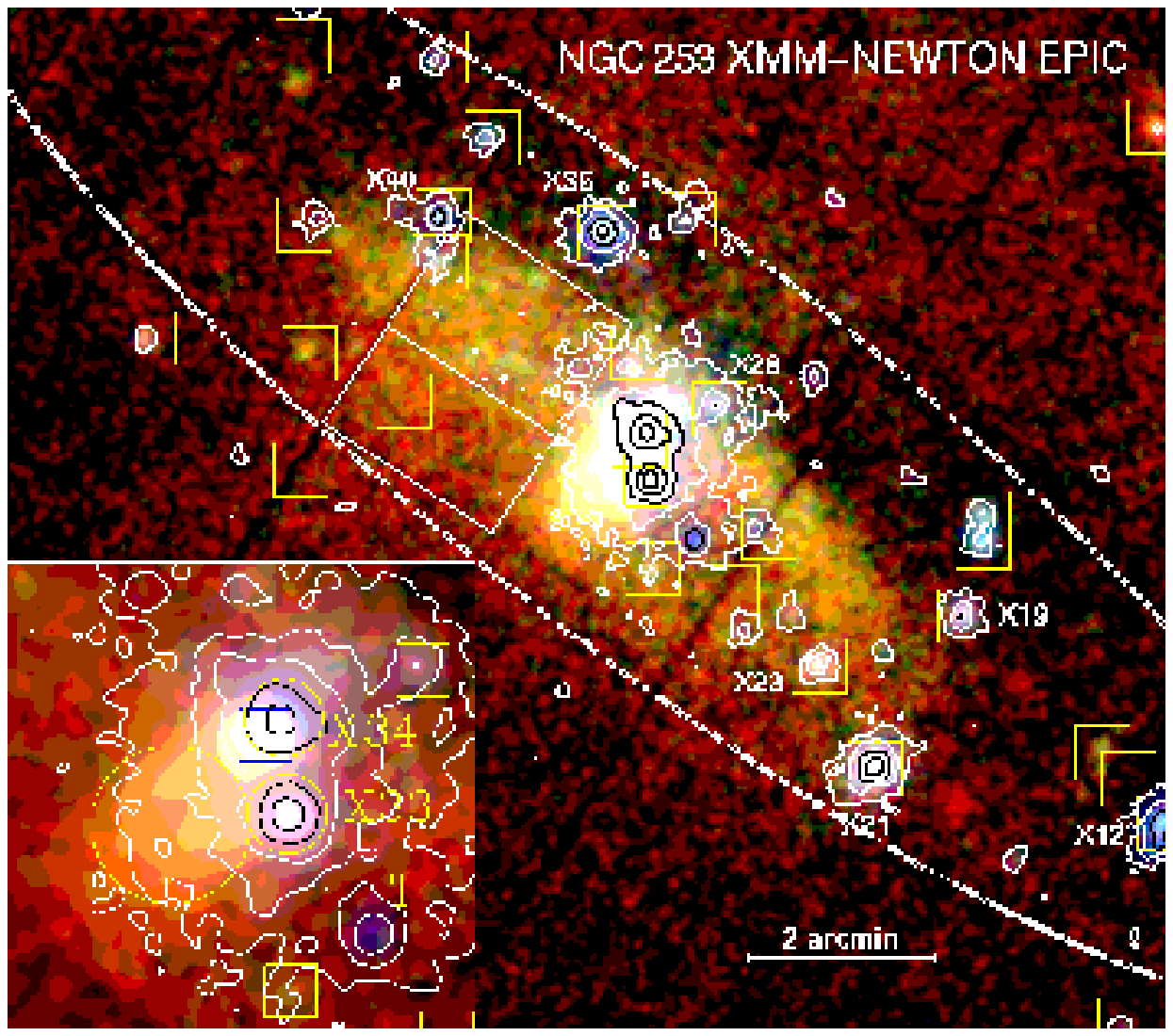}
    \includegraphics[width=8.9cm,bb=76 200 534 593,clip=]{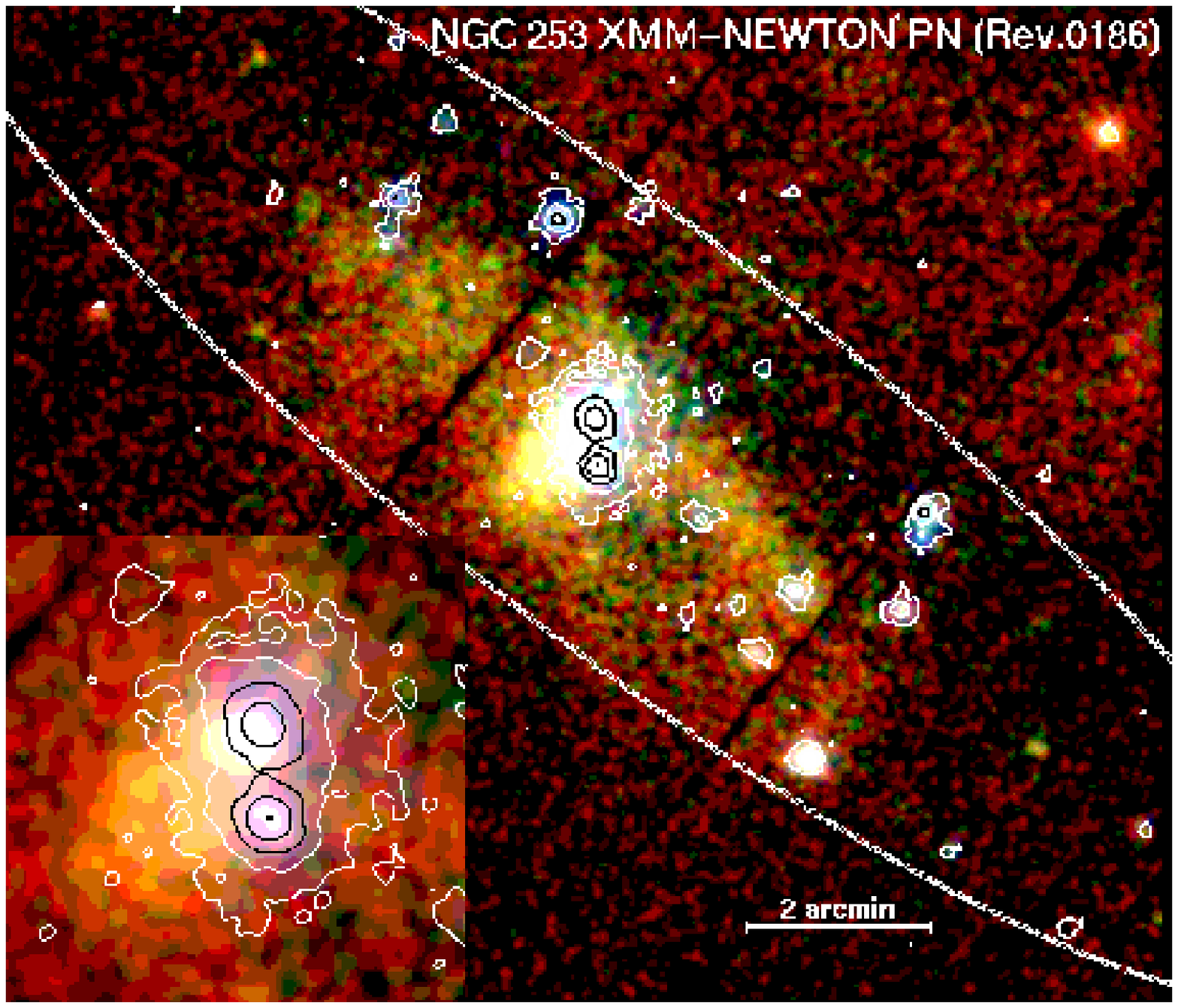}
  \end{center}
\caption{ XMM-Newton EPIC low background images of the \n253. 
The colours represent X-ray intensities in different energy
bands (red: 0.2--0.5 keV, green: 0.5--0.9 keV, blue: 0.9--2.0 keV), while the 
2--10 keV emission  is shown superimposed as contours.
The white ellipse indicates the optical $D_{25}$ diameter
of the galaxy: 
({\bf Left}) PV observation from July 2000 (see \protect\cite{wpietsch-E3:pie01}),
combining all EPIC instruments.
ROSAT sources from  \protect\cite*{wpietsch-E3:vog99} are indicated 
by squares. Spectral extraction regions are indicated.
({\bf Right}) SSC GT observation from December 2000, using only EPIC PN}  
\label{wpietsch-E3_fig7}
\end{figure*}

Figure~\ref{wpietsch-E3_fig7} allows us to compare EPIC images of the inner 
region of \n253\ from the PV and GT observation. 
The images are colour coded according to the
X-ray intensities in different bands with hard band contours superimposed. The
EPIC data clearly reveal more point sources than the deep ROSAT observations 
(squares on PV image give sources from \cite*{wpietsch-E3:vog99}) 
and allow the mapping of diffuse emission in the
disk. The underlying surface brightness at an energy of $\sim1$ keV increases by
factors of $\sim10$ between disk, plume and extended nucleus.  

All bright persistent sources known from the ROSAT observations are also
detected by \xmm. Some of the sources show brightness variations by factors of
several compared to the ROSAT luminosities. In addition we detect at least 3 
transient sources. The ROSAT source X12 was already classified as a transient by 
\cite*{wpietsch-E3:vog99} as it was only detected in the ROSAT observations in June
1992. X12 was again detected in the July 2000 \xmm\ observations, 
however was not present in December 2000. Therefore, X12 has to be classified as a 
recurrent transient. A second bright transient was detected $\sim70$\arcsec\
SSW of the \n253\ nucleus in the July 2000 observation and already was visible
during the Chandra observation half a year earlier (\cite{wpietsch-E3:str00}).
It was neither detected by ROSAT nor in the \xmm\ GT observation in December
2000. A third transient is detected for the first time in the December 2000
\xmm\ GT observation $\sim45$\arcsec\ NW of X23. 

\begin{figure*}[ht]
 \begin{center}
    \includegraphics[width=6.8cm,angle=-90,clip=]{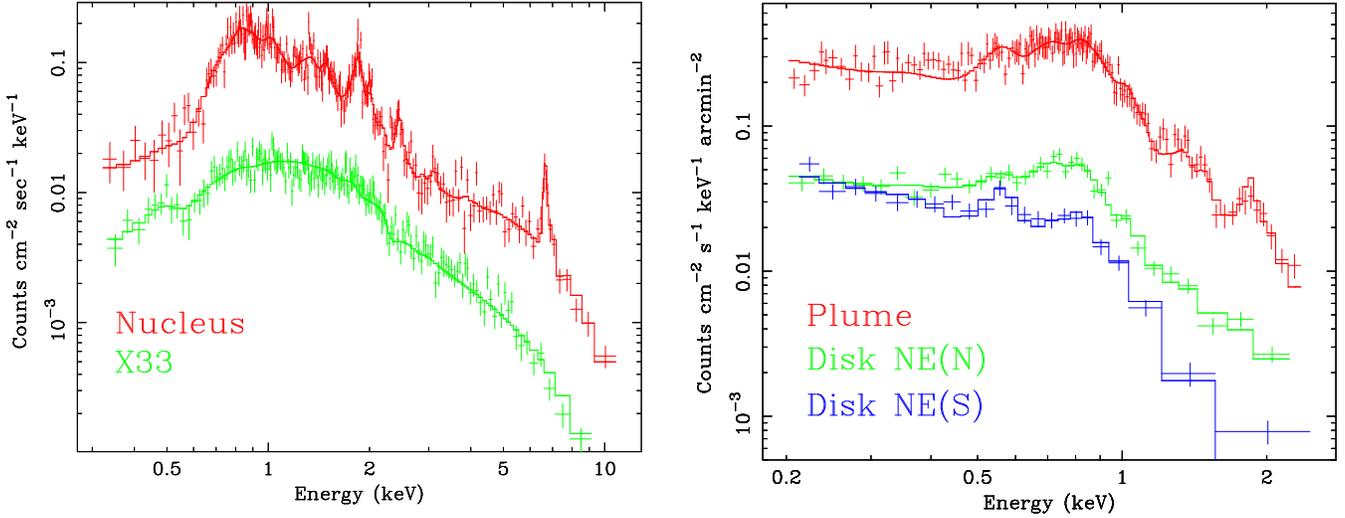}
    \includegraphics[width=6.8cm,angle=-90,clip=]{wpietsch-E3_fig9b.eps}
 \end{center}
\caption{ EPIC PN background subtracted spectra with spectal models indicated: 
({\bf Left}) spectra of the extended nucleus of \n253\ and the XRB \n253-X33.
({\bf Right}) spectra of the \n253\ SE X-ray plume and two areas of the disk 
NE of the nucleus}  
\label{wpietsch-E3_fig9}
\end{figure*}

The brightest point source, X33 just south of the nucleus, varies by a factor of
two during the PV observation and most likely is a black hole XRB as its
spectrum can be modeled by an absorbed disk blackbody
plus power law (see Fig.~\ref{wpietsch-E3_fig9}). Therefore the source is most 
likely a black hole XRB. 

\begin{figure}[ht]
  \begin{center}
    \includegraphics[width=8.9cm,bb=65 70 520 520,clip=]{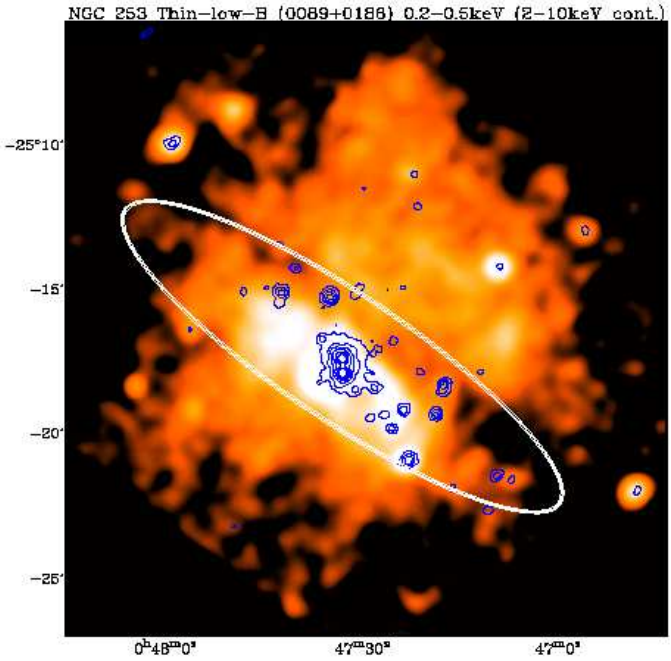}
  \end{center}
\caption{Smoothed EPIC PN thin filter low background 0.2--0.5 keV image of
the \n253\ field combining PV and SSC GT exposures. 
2--10 keV emission is shown superimposed as blue contours.
The white ellipse indicates the optical $D_{25}$ diameter
of the galaxy. }  
\label{wpietsch-E3_fig8}
\end{figure}

\ein\ and ROSAT observations have revealed four diffuse X-ray emission 
components in \n253: the nuclear area, the plume, the galaxy disk, 
and especially the halo out to
projected distances of 9 kpc from the disk (\cite{wpietsch-E3:fab84}, 
\cite{wpietsch-E3:fab88}, and e.g. \cite{wpietsch-E3:pie00}). The 
component with the softest spectrum originated from the halo of \n253, 
which was fitted in the ROSAT
PSPC observations with a temperature of 
\linebreak[4]
$\sim0.15$~keV. While the slightly
harder emission from the NW halo was already detected with \ein, the
soft response and the low background performance of the ROSAT PSPC 
was needed to detect the
softer emission from the SE halo. It was therefore a challenge for the soft
energy response of \xmm\ if this halo emission could be detected.

The first three components are easily visible in the colour images 
(Fig.~\ref{wpietsch-E3_fig7}) and 
have been spectrally analyzed in some detail (see description of the PV 
results below). To search for the halo component, all EPIC PN low background
thin filter observations were combined and the halo emission can be
clearly detected in the 0.2--0.5 keV image. The spatial shape reflects that 
known from
the ROSAT observations (Fig.~\ref{wpietsch-E3_fig8}). A detailed spectral
analysis of the very extended halo component is in progress.

The EPIC images reveal unresolved emission from the inner disk that is harder
along the inner spiral arms. EPIC PN spectra extracted NE of the nucleus,
selecting areas of harder emission close to the major axis (N) and
softer emission adjacent to the S (Fig.~\ref{wpietsch-E3_fig8}).
While the spectra below 0.5 keV are very similar (Fig.~\ref{wpietsch-E3_fig9}),
the NE(N) shows additional emission extending to energies of 
\linebreak[3]
$\sim2$~keV. They
both indicate emission lines from O\,{\sc vii} and Fe\,{\sc xvii} pointing at
major hot plasma origin. The spectra have therefore been modeled assuming ISM
components of solar abundance with two temperatures, shining through the ISM,
i.e. correcting for the Galactic foreground and adding additional absorption
within \n253. 
For the NE(N) spectrum a power law component had to be added which
may describe contributions from unresolved point sources.
The derived lower and higher temperatures (0.13~keV and 0.5~keV) agree in both
areas within the errors, and the cooler component does not need additional
absorption 
\linebreak[3]
with\-in \n253, indicating that the emission originates from the halo
above the disk. These temperatures, derived over small areas, agree with the
ROSAT results for the entire disk.
 
\begin{figure}[ht]
  \begin{center}
    \includegraphics[bb=-59 116 630 630,width=7.8cm,clip=]{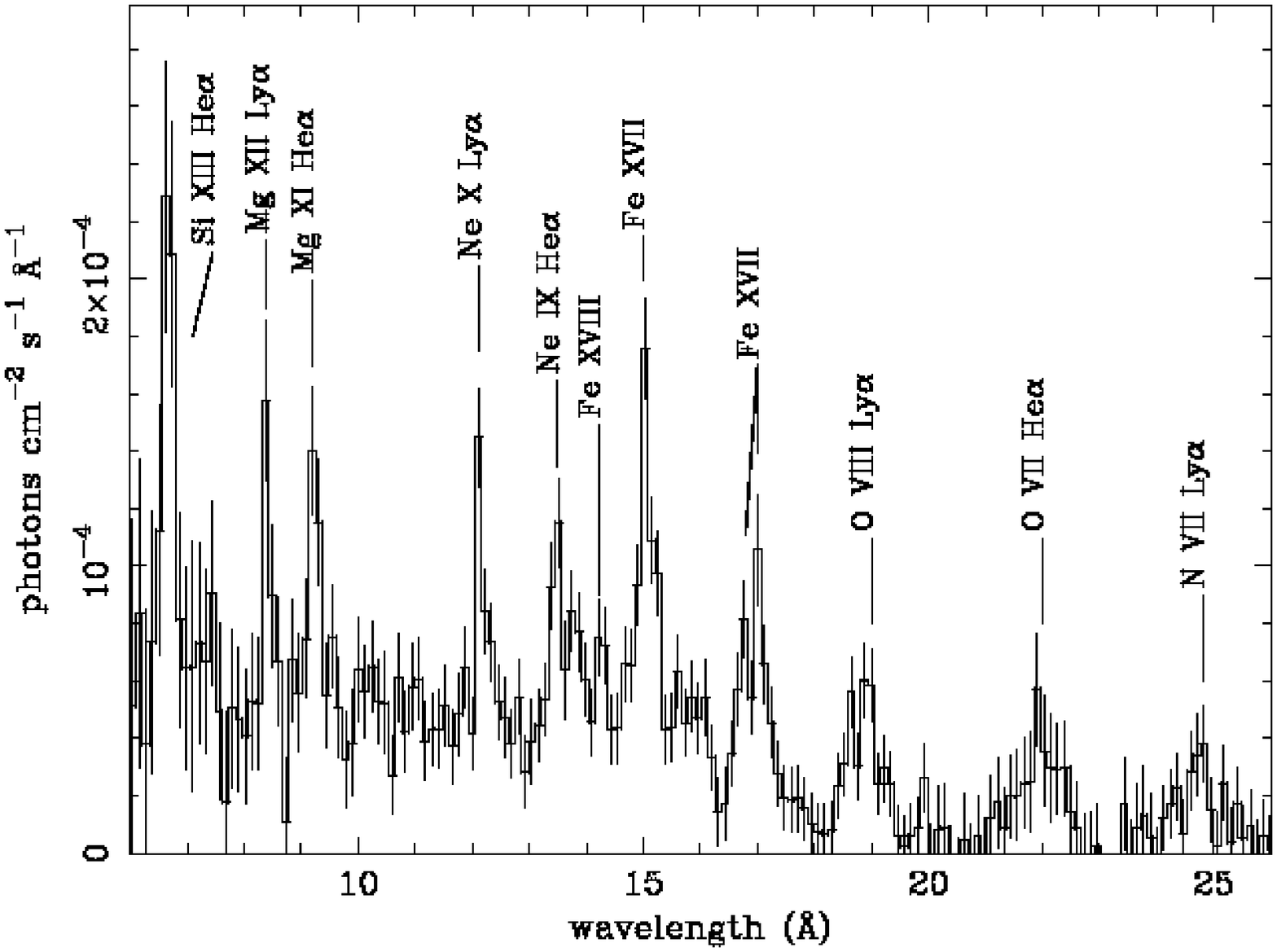}
  \end{center}
\caption{RGS spectrum of the bright nuclear area of \n253\ covering nucleus and
plume. Bright emission lines are identified. }  
\label{wpietsch-E3_fig10}
\end{figure}

The nuclear region of \n253 is bright enough for a detailed study with both RGS
and EPIC. The RGS spectrum (Fig.~\ref{wpietsch-E3_fig10}) is dominated by
emission lines of hydrogenic and heliumlike charge states of the abundant low Z
elements (N, O, Ne, Mg, Si) and the neonlike and fluorinelike charge states of
Fe. With the help of EPIC, one can further localize the RGS emission components
in the nuclear area. The strength of the Fe L lines relative to the K-shell
lines suggests that collisional ionization is the dominant soft X-ray emission
mechanism. The inferred temperatures range from 0.3 to 1.5 keV and the weakness
of the longer wavelength lines suggests significant absorption. The general
appearance of the RGS spectrum is reminiscent of the spectrum of intermediate
age SNR gas, as might be expected for a starburst nucleus and the interaction of
the outflowing wind with the cooler gas of the ISM in the plume.
In contrast to RGS, the EPIC instruments can spatially resolve the nuclear 
region in emission components from the unresolved nuclear source 
(X34) and the plume. 

No significant emission from X34 can be seen below 0.5 keV. As we go
to higher energies, the centroid of the X34 emission shifts towards the 
position of the radio nucleus indicating that only at energies above 4 keV 
emission from the nuclear starburst will dominate the X-ray spectrum. The
spectrum can be modeled using thin thermal plasma components of solar abundance
with corresponding absorption increasing with the temperature of the plasma
component, and an added power law component with the lowest absorption value.
Three temperature components give an acceptable fit with $N_{\rm H} = (0.34,
1.78, 13.2)
\linebreak[4]
\times 10^{22}$~\cm-2, photoindex of 1.0, and temperatures of (0.56,
0.92, 6.3)~keV, respectively. A model with 0.7 solar abundance for all components
and similar absorption and temperature values gives an equally acceptable fit
and does not need the power law component (Fig.~\ref{wpietsch-E3_fig7}). 
Note, that within the nuclear
spectrum no evidence for a significant AGN contribution is detected (which
would require a 
\linebreak[4]
highly absorbed non-thermal component) and the very hard
component can therefore be attributed purely to the starburst nucleus. 
Similar high-temperature plasma and Fe K lines have been found in young type Ib
and type IIa SNRs. If one assumes that such SNRs are responsible for the X-ray 
emission of the starburst nucleus and that these SNRs would be strong iron line
emitters for $\sim5000$ yr, one can estimate that 1000 SNRs would be needed and
the supernova rate of the \n253\ starburst should be 0.2~yr$^{-1}$.

The EPIC images (Fig.~\ref{wpietsch-E3_fig7}) trace the bright X-ray
plume emission of \n253\ out to 1\farcm75 (1300 pc projected distance) from the
nucleus along the minor axis of the galaxy disk to the SE into the galaxy halo,
much further than possible with earlier observations.
EPIC PN images for the
different ionization state ions seen with the RGS reveal that the
limb-brightening of the plume is mostly seen in higher ionization emission
lines, while in the lower ionization lines, and below 0.5 keV, the plume is more
homogeneously structured. The plume spectrum (Fig.~\ref{wpietsch-E3_fig9}) 
can be modeled by a three
temperature thermal plasma containing the two low temperature nuclear components
(though less absorbed) plus an unabsorbed 0.15 keV component similar to the disk
spectra. These new findings  point to new interpretations as to the make up of
the starburst-driven outflow.

\section{Summary}
The first observations of nearby late type galaxies have demonstrated the superb 
performance of the XMM-New\-ton instruments. However, sometimes high instrument 
\linebreak[4]
background causes problems by strongly reducing the useful observation time 
below the times derived in the feasibility calculations and scheduled for
observation. Another problem for homogeneous galaxy observations is imposed 
by the very stable \xmm\ attitude control system which does not 
smear out small scale detector structures  like CCD gaps or bad pixels. One
therefore might argue for the introduction of a satellite dithering mode. 

Besides these small obstacles, the big advantages of \xmm\ for nearby galaxy studies compared to earlier observatories 
are the
\begin{itemize} 
\item good spatial resolution and high collection power for EPIC and RGS spectra
\item high detection potential for XRBs, X-ray pulsars, SNRs, CVs
\item detectability of ISM and especially soft halo emission
($kT \approx 0.15$ keV)
in low background EPIC PN thin filter observations
\item advantage of simultaneous multi-instrument data collection 
\end{itemize}

The first \xmm\ observations promise a wealth of exciting 
results when new \xmm\ nearby galaxy observations are performed and analyzed.
The interpretation will certainly benefit when high spatial resolution 
\chandra\ observations of the galaxies are combined with the collecting power 
of \xmm. 
 
\begin{acknowledgements}
I thank my collaborators for their contributions to the work described above.
The \xmm\ project is an ESA Science Mission with instruments and
contributions directly funded by ESA Member states and the USA (NASA). 
    The \xmm\ project is supported by the Bundesministerium f\"{u}r
    Bildung und Forschung / Deutsches Zentrum f\"{u}r Luft- und Raumfahrt 
    (BMBF/DLR), the Max-Planck Society and the Heidenhain-Stiftung.

\end{acknowledgements}


\begin{thebibliography}{}

\bibitem[\protect\astroncite{Behar et~al.}{2001}]{wpietsch-E3:beh01}
Behar, E., Rasmussen, A.P., Griffiths, R.G. et al. \ 2001, A\&A 365, L242

\bibitem[\protect\astroncite{Dennerl et~al.}{2001}]{wpietsch-E3:den01}
Dennerl, K., Haberl, F., Aschenbach, B. et al. \ 2001, A\&A 365, L202

\bibitem[\protect\astroncite{Dubus et~al.}{1999}]{wpietsch-E3:dub99}
Dubus, G., Charles, P.A., Long, K.S., Hakala, P.J., Kuulkers, E., 1999,
MNRAS 302, 731

\bibitem[\protect\astroncite{Ehle et~al.}{2001}]{wpietsch-E3:ehl01}
Ehle, M., Pietsch, W., Haberl, F., 2001, in: New Century of X-ray Astronomy,
eds. H. Inoue \& H. Kunieda, ASP Conference Series 251, 300

\bibitem[\protect\astroncite{Fabbiano}{1988}]{wpietsch-E3:fab88}
Fabbiano, G., 1988, ApJ 330, 672

\bibitem[\protect\astroncite{Fabbiano \& Trinchieri}{1984}]{wpietsch-E3:fab84}
Fabbiano, G., Trinchieri, G., 1984, ApJ 286, 491

\bibitem[\protect\astroncite{Haberl \& Pietsch}{1999}]{wpietsch-E3:hab99}
Haberl, F., Pietsch, W., 1999, A\&AS 139, 277

\bibitem[\protect\astroncite{Haberl et~al.}{2000}]{wpietsch-E3:hab00}
Haberl, F., Filipovi\'c, M.D., Pietsch, W., Kahabka, P., 2000, A\&AS 142, 41

\bibitem[\protect\astroncite{Haberl \& Pietsch}{2001}]{wpietsch-E3:hab01}
Haberl, F., Pietsch, W., 2001, A\&A 373, 438

\bibitem[\protect\astroncite{Haberl et~al.}{2001}]{wpietsch-E3:ha101}
Haberl, F., Dennerl, K., Filipovi\'c, M.D., Aschenbach, B., Pietsch, W., 
Tr\"umper, J., 2001, A\&A 365, L208

\bibitem[\protect\astroncite{Osborne et~al.}{2001}]{wpietsch-E3:osb01}
Osborne, J.P., Borozdin, K.N., Trudolyubov, S.P. et al. \  2001, A\&A 378, 800

\bibitem[\protect\astroncite{Paerels et~al.}{2001}]{wpietsch-E3:pae01}
Paerels, F., Rasmussen, A.P., Hartmann, H.W. et al.\ 2001, A\&A 365, L303

\bibitem[\protect\astroncite{Pietsch et~al.}{1998}]{wpietsch-E3:pie98}
Pietsch, W., Trinchieri, G., Vogler, A., 1998, A\&A 340, 351

\bibitem[\protect\astroncite{Pietsch et~al.}{2000a}]{wpietsch-E3:pi100}
Pietsch, W., Vogler, A., Klein, U., Zinnecker, H., 2000a, A\&A 360, 24

\bibitem[\protect\astroncite{Pietsch et~al.}{2000b}]{wpietsch-E3:pie00}
Pietsch, W., Haberl, F., Ehle, M., Trinchieri, G., Vogler, A., 2000b, AAS,
HEAD meeting 32, 21.04

\bibitem[\protect\astroncite{Pietsch et~al.}{2001}]{wpietsch-E3:pie01}
Pietsch, W., Roberts, T.P., Sako, M. et al.\  2001, A\&A 365, L174

\bibitem[\protect\astroncite{Rasmussen et~al.}{2001}]{wpietsch-E3:ras01}
Rasmussen, A.P., Behar, E., Kahn, M.S., den Herder, J.W., van der Heyden, K.,
2001, A\&A 365, L231

\bibitem[\protect\astroncite{Sasaki et~al.}{2000a}]{wpietsch-E3:sas00}
Sasaki, M., Haberl, F., Pietsch, W., 2000a, A\&AS 143, 391

\bibitem[\protect\astroncite{Sasaki et~al.}{2000b}]{wpietsch-E3:sa100}
Sasaki, M., Haberl, F., Pietsch, W., 2000b, A\&AS 147, 75

\bibitem[\protect\astroncite{Sasaki et~al.}{2001a}]{wpietsch-E3:sa101}
Sasaki, M., Stadlbauer, T.F.X., Haberl, F., Filipovi\'c, M.D., Bennie, P.J.,
2001a, A\&A 365, L237

\bibitem[\protect\astroncite{Sasaki et~al.}{2001b}]{wpietsch-E3:sas01}
Sasaki, M., Haberl, F., Keller, S., Pietsch, W., 2001b, A\&A 369, L29

\bibitem[\protect\astroncite{Shirey et~al.}{2001}]{wpietsch-E3:shi01}
Shirey, R., Soria, R., Borozdin, K. et al. \  2001, A\&A 365, L195

\bibitem[\protect\astroncite{Soria et~al.}{2001}]{wpietsch-E3:sor01}
Soria, R., Wu, K., Page, M.J., Sakelliou, I., 2001, A\&A 365, L273

\bibitem[\protect\astroncite{Strickland et~al.}{2000}]{wpietsch-E3:str00}
Strickland, D.K., Heckman, T.M., Weaver, K.A., Dahlem, M., 2000, AJ 120, 2965

\bibitem[\protect\astroncite{van der Heyden et~al.}{2001}]{wpietsch-E3:van01}
van der Heyden, K.J., Paerels, F., Cottam, J., Kaastra, J.S., 
Branduardi-Raymont, G., 2001, A\&A 365, L254

\bibitem[\protect\astroncite{Vogler \& Pietsch}{1999}]{wpietsch-E3:vog99}
Vogler, A., Pietsch, W., 1999, A\&A 342, 101

\bibitem[\protect\astroncite{Wu et~al.}{2001}]{wpietsch-E3:wu01}
Wu, K., Soria, R., Page, M.J., Sakelliou, I., Kahn, S.M., de Vries, C.P.,
2001, A\&A 365, L267

\end{thebibliography}
\end{document}